\documentclass[conference]{IEEEtran}
\IEEEoverridecommandlockouts
\usepackage{cite}
\usepackage{amsmath,amssymb,amsfonts}
\usepackage{algorithmic}
\usepackage{graphicx}
\usepackage{textcomp}
\usepackage{xcolor}
\usepackage{libertine}
\usepackage{xspace}
\usepackage{pifont}
\usepackage{comment}
\usepackage{verbatim}
\usepackage{algorithmic}
\usepackage{algorithm}
\usepackage{url}
\usepackage{listings}
\usepackage{xcolor}

\newcommand{\smartfx}{\mbox{\textsc{SmartFX}}\xspace}

\def\BibTeX{{\rm B\kern-.05em{\sc i\kern-.025em b}\kern-.08em
    T\kern-.1667em\lower.7ex\hbox{E}\kern-.125emX}}
\begin{document}

\title{Constraint-Based Inference of Heuristics for Foreign Exchange Trade Model Optimization}

\author{\IEEEauthorblockN{Nikolay Ivanov}
\IEEEauthorblockA{\textit{Dept. of Computer Science and Engineering} \\
\textit{Michigan State University}\\
East Lansing, MI, USA \\
ivanovn1@msu.edu}
\and
\IEEEauthorblockN{Qiben Yan}
\IEEEauthorblockA{\textit{Dept. of Computer Science and Engineering} \\
\textit{Michigan State University}\\
East Lansing, MI, USA \\
qyan@msu.edu}
}


\maketitle

\begin{abstract}

The Foreign Exchange (Forex) is a large decentralized market, on which trading analysis and algorithmic trading are popular. Research efforts have been focusing on proof of efficiency of certain technical indicators. We demonstrate, however, that the values of indicator functions are not reproducible and often reduce the number of trade opportunities, compared to price-action trading.

In this work, we develop two dataset-agnostic Forex trading heuristic templates with high rate of trading signals. In order to determine most optimal parameters for the given heuristic prototypes, we perform a machine learning simulation of 10 years of Forex price data over three low-margin instruments and 6 different OHLC granularities. As a result, we develop a specific and reproducible list of most optimal trade parameters found for each instrument-granularity pair, with 118 pips of average daily profit for the optimized configuration.
\end{abstract}


\section{Introduction}
The Foreign Exchange, also known as Forex, is a decentralized global market with trillions of dollars of daily turnover~\cite{britanicaforex}. Some unique features of the Forex market make it highly suitable for technical analysis and algorithmic trading~\cite{thompson2017time}. Technical indicator functions are popular heuristics used by Forex traders to identify trading opportunities. Multiple research efforts have been focusing on the proof of efficiency of certain existing indicator functions for fixed trade setups~\cite{yazdi2013technical,5358946, krishnan2009impact,yong2015technical,ozturk2016heuristic,yazdi2012technical,ni2009exchange}. However, as we demonstrate in Section~\ref{section:preevaluation}, the values of popular indicators are different for different data sets and brokers. This compromises the accuracy of machine learning of the historical Forex data and impedes the development of concrete recommendation for the traders. Moreover, we also demonstrate that the indicator heuristics reduce the overall count of trade opportunities, and thus reduce the final profit.

In this paper, we develop two dataset-agnostic price-action heuristic templates for trade signals on the Forex Exchange market: \emph{candle trend continuation} and \emph{candle trend reversal}. These heuristics are characterized by a very high rate of trading opportunities, compared to popular technical indicators. The ultimate goal of this work is to determine most optimal parameters for the given heuristic templates for each combination of Forex instruments and trade periods given in the input data set. In order to achieve that goal, we develop \smartfx, a machine learning tool which analyzes 10 years of historical Forex price data over three low-margin instruments and 6 different OHLC granularities, delivering a specific and reproducible list of most optimal trade parameters found for each input granularity of each instrument.

\subsection{Motivation}

Although the research and analysis of the Forex market is abound with original approaches, the practical application of these approaches is challenged by the ambiguity of results, unrealistic assumptions, or simply the lack of proper methodology for historical data analysis~\cite{rosenstreich2005forex,guides2017personality}.

Thus, the motivation of this work is to deliver a clear, specific, evidence-supported, realistic, and reproducible practical trading model with concrete parameters that can be applied by all tiers of Forex traders.

\section{Background}

\subsection{Definitions}

\paragraph{Foreign Exchange (Forex)} The Foreign Exchange is a global market in which national currencies and some commodities are exchanged~\cite{britanicaforex}.

\paragraph{Forex Instrument (Forex Pair)} Forex Instrument is a pair of equities, which are either
currencies (e.g. U.S. Dollar) or commodities (e.g., gold). For example, the pair EUR/USD is a Forex instrument for trading U.S. Dollars for Euros and vice versa. Some examples of Forex instruments are given in Table~\ref{table:instruments}. The taxonomy of Forex pairs includes several types, such as \emph{major pair} or \emph {cross pair}, but they are all traded in the same manner.

\paragraph{Pip} Pip is a small unit of price movement (change) in the Forex market. The value of one
pip is different depending on the instrument. For EUR/USD 1 pip = 0.0001,
for USD/JPY and EUR/JPY, 1 pip = 0.01.

\begin{table}[]
\caption{Examples of Forex instruments and their types.}
\label{table:instruments}
\begin{tabular}{|l|l|l|}
\hline
\textbf{Instrument} & \textbf{Description}               & \textbf{Type} \\ \hline
EUR/USD             & Euro/U.S. Dollar                   & major pair    \\ \hline
USD/JPY             & U.S. Dollar/Japanese Yen           & major pair    \\ \hline
EUR/JPY             & Euro/Japanese Yen                  & cross pair    \\ \hline
GBP/AUD             & Pound Sterling/Australian Dollar   & cross pair    \\ \hline
NZD/CAD             & New Zealand Dollar/Canadian Dollar & cross pair    \\ \hline
XAU/USD             & Gold/U.S. Dollar                   & commodity     \\ \hline
XAG/USD             & Silver/U.S. Dollar                 & commodity     \\ \hline
\end{tabular}
\end{table}

\paragraph{Long trade} The Forex market allows two trade directions: buying and selling. The buying trade is commonly called a \emph{long trade}.

\paragraph{Short trade} The Forex market allows to submit a selling trade even if there is no active buying trade; for a trader it may be perceived as betting on a downward price trend or as a reverse trade. Regardless of the trader's perception, the selling trade is commonly called a \emph{short trade}.

\subsection{Spread}
Many Forex brokers do not charge any direct per-trade commission. Instead, they establish a difference between the buying price (called the \emph{ask price}, denoted $P_a$) and the selling price (called \emph{bid price}, denoted $P_b$). Then the \emph{spread} of a broker $\beta$ for an instrument $i$, denoted $P_{\beta,i}$, is defined as follows\footnote{In some literature, the spread is represented as a negative value. We use the absolute value notation to emphasize the simple fact of the difference, regardless of its sign.}:
$$
S_{\beta,i} = |P_b - P_a|
$$

\subsection{Margin trading}
In most days when the market is open, the Forex prices experience only minor fluctuations, which would require traders to make very large deposits in order to gain a feasible profit. In order to make price movement more influential, most Forex brokers allow for \emph{margin trading}, in which the broker lends the client an amount of money $V_{m}$, called the \emph{margin}, which is $M$ times greater than the actual deposit $D$, thus:
$$
V_{m} = M \times D
$$

The value $M$, usually denoted $M:1$, e.g., 1:50, is called the \emph{leverage}. Most Forex brokers in the United States offer the 1:50 leverage.

\subsection{OHLC Candlestick Data}

The raw Forex price data is usually provided by a broker in the form of a series of \emph{ticks}. Each instrument has its own stream of ticks, and each tick includes a timestamp, bid price, ask price, and an optional trade volume data. Depending on the broker and its API, the period between ticks can vary between 50 ms to several seconds. The tick data is necessary for trading, but in most cases it is overwhelmingly excessive for a technical analysis. Thus, a more analysis-friendly cmopressed form, called \emph{OHLC candlestick data}, is commonly used in most markets, including Forex.

\begin{figure}[H]
    \centering
    \includegraphics[width=3.3in]{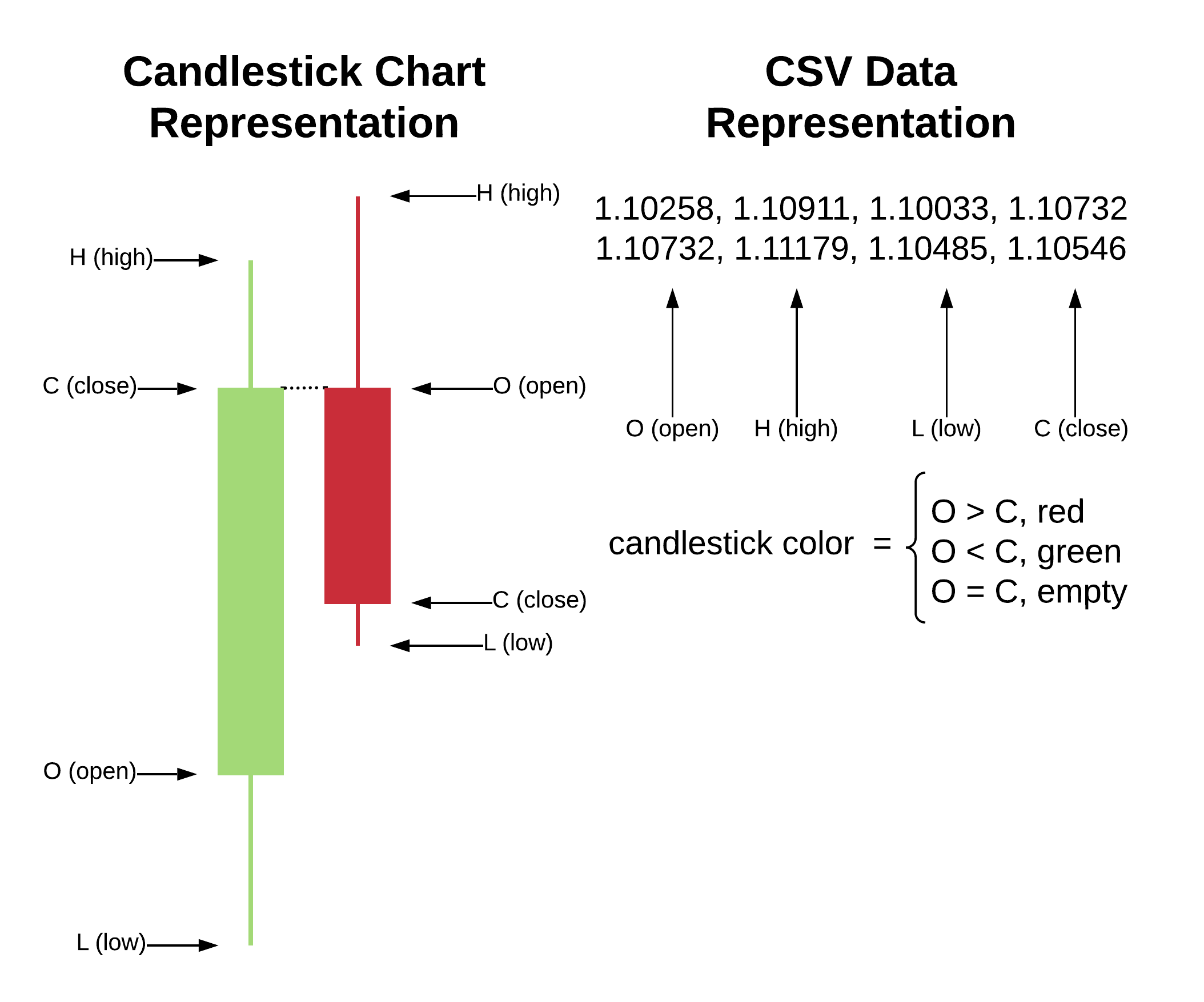}
    \caption{The OHLC candlestick data in chart and comma-separated value (CSV) representations.}
    \label{fig:candlesticks1}
\end{figure}

OHLC (Open, High, Low, Close) data is a set of points, called \emph{OHLC candles}, \emph{OHLC candlesticks}, or \emph{OHLC ticks}, summarizing the price movement during a fixed period of time using four parameters: 1) price at the beginning of the period (O); 2) maximum price during the period (H); 3) minimum price during the period (L); and 4) the price at the end of the period (C). Figure~\ref{fig:candlesticks1} demonstrates the chart and CSV representations of OHLC data. The period of time that one candle spans in a data set is called the \emph{OHLC granularity}. Common OHLC granularities for the Forex market are: M1 (1 minute), M2 (2 minutes), M5 (5 minutes), M10 (10 minutes), M15 (15 minutes), M30 (30 minutes), H1 (1 hour), H2 (2 hours), H4 (4 hours), H8 (8 hours), H12 (12 hours), D1 (one day), W1 (one week), and MN (one month).

\subsection{Forex Analysis}
The goal of a the Forex analysis is to predict the future movement of the price and identify trading opportunities. There are two major types of the market analysis: a) the \emph{fundamental analysis} and b) the \emph{technical analysis}. The fundamental analysis uses economic, political, and other news data to predict the future behavior of the market. The technical analysis uses present and past movement of the market price to infer the future move of the price, regardless of any outside (fundamental) events. Many traders also use various combinations of the fundamental and technical analyses. The technical analysis may be based on heuristic functions called \emph{technical indicators}, or it can be based solely on the past and present values of the price. The latter case is called the \emph{price-action}. \smartfx focuses on the pure price-action technical analysis only.

\section{Related Work}

    \subsection{Research}
Automated Forex exchange trading is a popular topic among peer-reviewed studies~\cite{de2012comparative,5358946,yazdi2013technical,8096360,wilson2010interday,pinto2012strategic}. In the last two decades, many studies have been conducted on different aspects of foreign exchange technical indicators~\cite{vezeris2019adturtle,5358946,yazdi2013technical,8096360,wilson2010interday,pinto2012strategic}. Many patents have been filed for automatic Forex trading systems~\cite{schoen2011automated,chait2007foreign,peterson2005foreign,ogg2005automated,owens2005automated}. Also, the Artificial Intelligence and Machine Learning has been used for price prediction of the Foreign Exchange market~\cite{983110,abraham2001intelligent,czekalski2015ann,lee2014hidden,yazdi2013technical,barbosa2010multi,shahbazi2016forex}.

However, the number of combinations of possible settings, as well as the abundance of pre-existing constraints for Forex automated trading is virtually infinite. As the number of constraints is growing, the necessity to employ the methods of Artificial Intelligence and Machine Learning becomes an urgent necessity because the space and time complexity of a brute-force trial exceeds storage and computational abilities of existing computers. To the best of my knowledge, none of the existing studies delivers a complete set of trading parameters for automated or semi-automated Forex trading with low equity and other parameters representing a non-institutional trader.
    
\subsection{Trading Heuristics}

Most popular heuristics for technical analysis are technical indicators~\cite{ozturk2016heuristic}. For example, one of the largest Forex brokers in the United States, GAIN Capital Group LLC~\cite{macerinskiene2015evidence}, provides 75 classes of technical indicators as part of its Web Trading Platform~\footnote{https://webtrader.forex.com/workspaces/default}.
    
\subsection{Retrospective Simulation}

The retrospective simulation of Forex trading is a software-based processing of historical Forex price data, using a given trade model and a set of heuristics. There is a number of software trade simulators available on the market, such as MetaTrader\footnote{https://www.metatrader4.com/en}, NinjaTrader\footnote{https://ninjatrader.com/Simulate}, Forex Tester\footnote{https://forextester.com/}, Soft4Fx\footnote{https://soft4fx.com/}, fxSimulator\footnote{https://fxsimulator.com/}, and others. All these tools allow to test a given strategy, automatically or manually, using a historical data. However, to the best of my knowledge, none of these tools is capable of \emph{calculating the most optimal strategy}, which is the main goal of \smartfx.
    
\subsection{Other Works}

There are multiple books written on the topic of Forex trading optimization. The "forex trading" search result at the Book Department of Amazon web store returns more than 6,000 results\footnote{https://www.amazon.com/s?k=forex+trading}. The same search request on a popular platform for online video courses, Udemy, returns 3,934 results\footnote{https://www.udemy.com/courses/search/?src=ukw\&q=forex\%20trading}.

\section{Pre-Evaluation}\label{section:preevaluation}

\subsection{Inconsistency of Technical Indicators}

Moving Average Convergence/Divergence (MACD) is one of the most popular technical indicators among Forex traders~\cite{yazdi2013technical}. However, different data sets and different trading tools reveal significant differences between the actual values of popular technical indicators, and this is easily demonstrable by comparing the charts of MACD for the same instrument with the same OHLC granularity within the same time frame, with time zone differences properly synchronized.

\begin{figure}[H]
    \centering
    \includegraphics[width=1.5in]{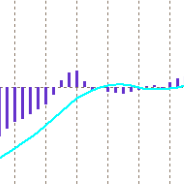}~~~
    \centering
    \includegraphics[width=1.5in]{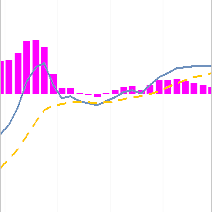}\\
    ~(a) Forex.com/MT4~~~~~~~~~(b) Dukascopy Demo/JForex\\
    \vspace{0.5cm}
    \centering
    ~\includegraphics[width=1.5in]{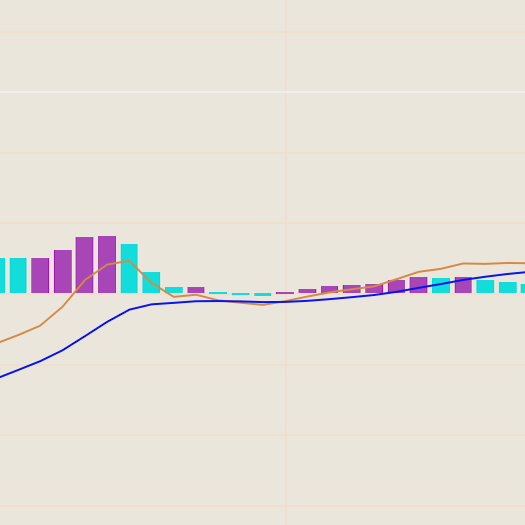}~~~
    \centering
    \includegraphics[width=1.5in]{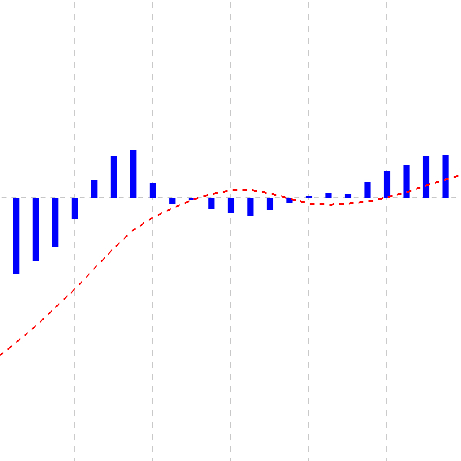}\\
    ~~~~~~(c) Oanda/iOS~~~~~~~~~(d) Metaquotes Demo/MT5 iOS\\

    \caption{MACD charts produced by different trading tools, which use different data sets for the same period of time (November 28, 2019 00:00---November 28, 2019 23:59, EST, adjusted for differences in time zones) of the same Forex instrument (EUR/USD) with the same OHLC granularity (H1).}
    \label{fig:preeval}
\end{figure}

For this experiment, we observe four charts of the MACD technical indicator, produced by different trade tools with the same parameters (fast MA: 12, slow MA: 26, length: 9) generated for 24 OHLC H1 candlesticks representing the Forex price during November 28, 2019 EST, accounting for the differences in time zones\footnote{Two out of four data sets used the GMT time zone, which was shifted to match the EST time zone.}. As we can see in Figure~\ref{fig:preeval}, there is a significant differences between the MACD indicators with the same parameters over the same time period. One reason for such a discrepancy may be the fact that November 28, 2019 was the Thanksgiving Day in the United States, during which the trading activity is very low, whereas in most other countries it was just a regular business day, resulting in Forex activity difference in brokers from different regions (which has some effect upon the price). Regardless of the reason, the experiment demonstrated that \emph{there was at least one day, during which MACD showed significantly different results for different brokers}.

\subsection{Trade Opportunities Rate Reduction}

In this work, we propose a price-action trade model, in which every OHLC candle, upon its closure time, delivers a new trade signal. For example, for the H1 OHLC granularity, the number of trading signals per day is 24. In other words, in \smartfx the signal period $T_{SmartFX}$ is constant and equal to the granularity $G$:
$$
T_{SmartFX} = G
$$

When the trade signals come from traditional technical indicators, the trade signal period $T_{ind}$ is variable, and presumably many times larger than $G$. In order to demonstrate this, we conducted a short experiment, in which we counted the number of possible configurations for MACD and RSI technical indicators that can be treated as signals over one month of historical Forex data (November 2019, EUR/USD) with OHLC granularity H1. Table~\ref{table:macdrsi-experiment} shows the results of this experiment.

\begin{table}[H]
\caption{Number of trade signal signatures per month produced by MACD, RSI, and \smartfx  for EUR/USD pair with OHLC granularity H1, counted during November 2019. For \smartfx, se assume that there are 20 full days when the Forex market is open during the month.}
\label{table:macdrsi-experiment}
\begin{tabular}{|c|l|c|}
\hline
\multicolumn{1}{|l|}{\textbf{Heuristic}} & \textbf{Heuristic Parameters}                                                 & \multicolumn{1}{l|}{\textbf{\begin{tabular}[c]{@{}l@{}}Number of Signal\\ Signatures During \\
November, 2019 \\
\end{tabular}}}\\ \hline
MACD                                     & \begin{tabular}[c]{@{}l@{}}Fast MA: 12\\ Slow MA: 26\\ Length: 9\end{tabular} & 55                                                                                                             \\ \hline
RSI                                      & Length: 14                                                                    & 64                                                                                                             \\ \hline
\smartfx                                  & N/A                                                                           & 480                                                                                                            \\ \hline
\end{tabular}
\end{table}

Let us show that on average the technical indicators produce less trade signals than \smartfx. Suppose that there is a set of parameters $\Upsilon$ for a traditional technical indicator $I$, which is twice as more successful than \smartfx (with the same take-profit/stop-loss settings), and which also produces twice as more trade signals than RSI-14. Then, if the average proportion of successful trades for \smartfx is $N_{SmartFX} (0 \le N_{SmartFX} \le 1)$, then the number of successful trades per month for \smartfx will be as follows:

$$
ST_{SmartFX} = 480 \times N_{SmartFX}
$$

Whereas the number of successful trades for the hypothetical indicator will be as follows:

$$
ST_{I(\Upsilon)} = 2 \times (2 \times 64 \times N_{SmartFX}) = 256 \times N_{SmartFX} 
$$

As it is seen from the experiment, even with unrealistically optimistic assumptions, $ST_{SmartFX} > ST_{I(\Upsilon)}$. And therefore it is fair to conclude that even if the traditional technical indicators have higher rate of successful signals, the decreased overall number of trading opportunities yields a lower overall profit compared to per-candle price action.

\section{SmartFX Design}

\begin{figure}[]
    \centering
    \includegraphics[width=3.2in]{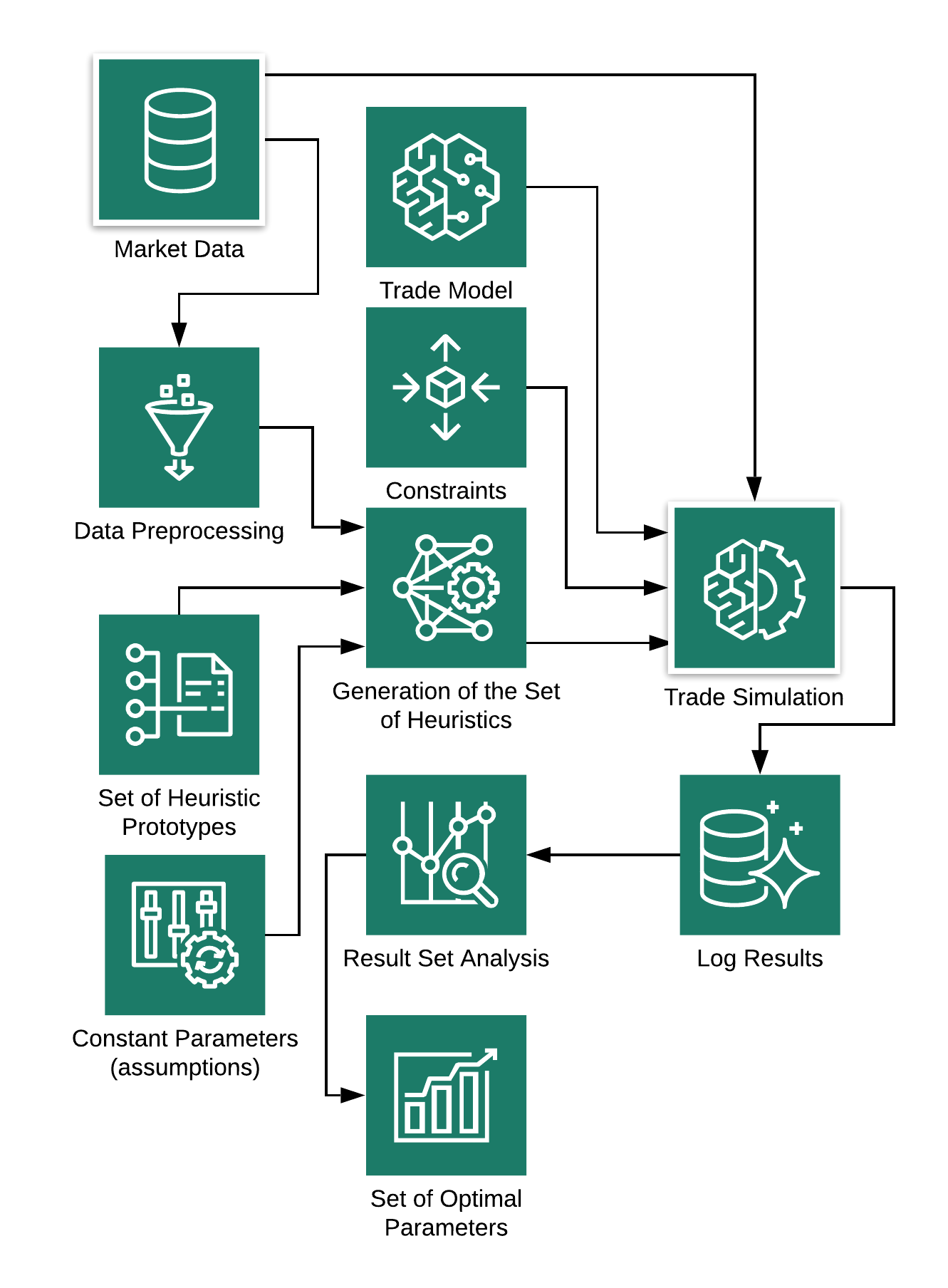}
    \caption{An overview of the \smartfx design. The market data, which consists of several subsets, is pre-processed for performance optimization by generating an auxiliary hash maps for the data subsets. The pre-processed data is used by the generator of heuristics, along with the set of heuristic prototypes and constant parameters (e.g., broker's spread). The generated heuristics are fed to the multi-threaded trade simulation learning algorithm, along with a set of constraints (representing the trader), and the trade model (representing the rules of the Forex market). The results of the simulation are being analyzed to produce a set of the most optimal trading parameters.}
    \label{fig:smartfxdesign}
\end{figure}

Figure~\ref{fig:smartfxdesign} demonstrates the basic design of \smartfx. The market data, which is described in detail in Section~\ref{section:dataset}, is being pre-processed to facilitate the performance of the simulation. Specifically, the market data is being saved twice in two different data structures: hash table and serial list. This allows to reduce the processing time because the latter is the system's performance bottleneck.

The pre-processed market data goes to the generator of the set of heuristics, along with the set of heuristic prototypes and constant parameters (or assumptions). The two heuristic prototypes are presented in Algorithm~\ref{alg:alg1} and Algorithm~\ref{alg:alg2}. The constant parameters include: maximum total loss allowed in pips, maximum take profit, maximum stop loss, take profit and stop loss steps, and the broker's spread.

The fist heuristic prototype, called the \emph{trend continuation heuristic}, denoted as $h_{1}$, simply signals a long trade at the end of a ``green'' candle period, and it signals the short trade otherwise. The second heuristic prototype, called the \emph{trend reversal heuristic}, denoted as $h_{2}$, signals a short trade at the end of a ``green'' candle period, and it signals a long trade otherwise. It is important to note that in order to avoid risks, only one trade at a time can be placed in \smartfx model.

 \begin{algorithm}[]
 \caption{Trend Continuation Heuristic $h_1$}
 \label{alg:alg1}
 \begin{algorithmic}[1]
 \renewcommand{\algorithmicrequire}{\textbf{INPUT:}}
 \renewcommand{\algorithmicensure}{\textbf{OUTPUT:}}
 \REQUIRE $O, H, L, C$
 \ENSURE  trade decision
  \IF {no current trades}
  \IF {$C > O$}
  \RETURN long trade
  \ELSIF {$C < 0$}
  \RETURN short trade
  \ELSE
  \RETURN no trade
  \ENDIF
  \ENDIF
 \end{algorithmic}
 \end{algorithm}

\begin{algorithm}[]
 \caption{Trend Reversal Heuristic $h_2$}
 \label{alg:alg2}
 \begin{algorithmic}[1]
 \renewcommand{\algorithmicrequire}{\textbf{INPUT:}}
 \renewcommand{\algorithmicensure}{\textbf{OUTPUT:}}
 \REQUIRE $O, H, L, C$
 \ENSURE  trade decision
 \IF {no current trades}
  \IF {$C < O$}
  \RETURN long trade
  \ELSIF {$C > 0$}
  \RETURN short trade
  \ELSE
  \RETURN no trade
  \ENDIF
  \ENDIF
 \RETURN no trade
 \end{algorithmic}
 \end{algorithm}

When the set of heuristics is generated, they go to the trade simulation algorithm along with the trade model (which represents the rules of the Forex market) and the set of constraints (which represents the parameters of a Forex trader). The trade model, constraints, and the set of heuristics are the same for each subset of the market data. The subsets of market data are formed by combining all the given instruments with all the given OHLC granularities. Each subset of data can be independently processed by the trade simulation algorithm, thus this algorithm can be run in a multi-threaded environment.

Each simulation thread maintains its own log of results, which is analyzed to determine the set of optimal parameters. Then the sets of all optimal parameters for each subset of the market data are processed manually in a spreadsheet software.

\section{SmartFX Implementation}

\smartfx is a multi-threaded open-source tool written in Java, totalling 859 lines of code. The value of spread is fixed and set to 2.0 pips. The software performs data preprocessing and generation of heuristics in a linear fashion. However, the most computation-consuming part, the trade simulation learning, is performed in a dedicated thread for each subset of the market data.

The range of the values of the stop loss (SL) parameter is set to [3,50]. The same range is used for the take profit (TP) parameter, accounting for 2,209 distinct $(SL,TP)$ tuples. We stochastically determined that the largest ranges are not likely to yield any combinations with better efficiency. Moreover, higher values of TP reduce the number of trades, whereas larger value of SL increases the risk of unexpected large loss of profit.

\section{Data Set}\label{section:dataset}

We use 18 subsets of historical Forex data from the Dukascopy Swiss Banking Group data set\footnote{https://www.dukascopy.com/swiss/english/marketwatch/historical/}. The subsets are tuples $(\eta, G)$, where $\eta$ is a Forex instrument, and $G$ is a granularity, such that:
$$
\eta \in \{EUR/USD, USD/JPY, EUR/JPY\}
$$
$$
G \in \{H1, H2, H4, H8, H12, D1\}
$$

Each subset is stored in a comma-separated value (CSV) data file. Each CSV entry includes the OHLC candlestick data. The entries also include the trade volume data, but this parameter is ignored by \smartfx.

\smartfx records the following parameters for each of the 36 runs (18 subsets of market data processed with each of the two heuristic prototypes) during the learning simulation: a) total profit in pips; b) total loss in pips; c) heuristic used; d) take profit (TP) policy simulated; e) stop loss (SL) policy simulated; f) number of long trades; g) number of short trades; h) Forex instrument; and i) granularity of OHLC data.

\section{Evaluation}

\subsection{Setup and Run}

We used a Dell PowerEdge T640 server with two Intel Xeon Gold 5218 processors and 200 GB of RAM, running Ubuntu 18.04.03 LTS. The \smartfx tool was executed in two separate processes, each running 18 threads (36 threads total). The first process was resposible for the heuristic $h_1$, and the second process was responsible for the heuristic $h_2$. The choice to run the tool in two multi-threaded processes was determined by implementation convenience, and was not intended to achieve any performance improvement. The length of the entire cycle for all data subsets took about two hours.

\subsection{Results}

The raw results of \smartfx execution are recorded in the Appendices, and below are the most essential summaries and explanations.

\subsubsection{Worst Performers}

Table~\ref{table:worstperformers} presents the four worst 10-year balances. It is important to note that these cases are worst compared to other cases, but they are best cases locally. For example, the first row indicates that the best combination of take profit and stop loss values for the EUR/JPY instrument for OHLC H2 granularity and trade continuation heuristic could maximize the ending balance only to the value of -75,380 pips. As we can see, all the four worst performers use the trend continuation heuristic prototype. The performance of the two heuristic prototypes is discussed later in this section.

\begin{table}[]
\caption{Four simulations that resulted in lower 10-year ending balances.}
\label{table:worstperformers}
\begin{tabular}{|l|l|l|l|}
\hline
\textbf{Instrument} & \textbf{Period} & \textbf{Heuristic} & \textbf{10-year Balance (pips)} \\ \hline
EUR/JPY             & H2              & $h_1$                 & -75,380                  \\ \hline
EUR/JPY             & H4              & $h_1$                 & -41,338                  \\ \hline
EUR/USD             & H4              & $h_1$                 & -39,953                  \\ \hline
EUR/USD             & H2              & $h_1$                 & -29,397                  \\ \hline
\end{tabular}
\end{table}

\subsubsection{Best Performers}
\label{section:bestperformers}

The best four performers, shown in Table~\ref{table:bestperformers}, all use the trend reversal heuristic prototype. Also, the best four performance have finer OHLC granularities than the four worst performing cases. As we can see from the table, the \smartfx model can deliver on average up to 118 pips of profit. The performance of the two heuristics is discussed in later in this section.

\begin{table}[]
\caption{Four simulations that resulted in best 10-year ending balances.}
\label{table:bestperformers}
\begin{tabular}{|l|l|l|l|l|}
\hline
\textbf{Instrument} & \textbf{Period} & \textbf{Heuristic} & \textbf{\begin{tabular}[c]{@{}l@{}}10-year\\ Balance (pips)\end{tabular}} & \textbf{\begin{tabular}[c]{@{}l@{}}Avg. pips\\ per day\end{tabular}} \\ \hline
EUR/JPY             & H1              & h2                 & 432,192                                                                   & 118.31                                                               \\ \hline
EUR/USD             & H1              & h2                 & 348,611                                                                   & 95.43                                                                \\ \hline
EUR/JPY             & H2              & h2                 & 302,022                                                                   & 82.68                                                                \\ \hline
USD/JPY             & H1              & h2                 & 293,121                                                                   & 80.24                                                                \\ \hline
\end{tabular}
\end{table}

\subsubsection{Number of Trades}

Table~\ref{table:numberoftrades} summarizes the number of trades statistics among the top 4 performers discussed before in this section. We are interested in the number of trades only for top performers because they are the ones to be used in real trades. However, the information about the number of trades for the remaining cases is available in Appendix~\ref{appendix:numberoftrades}.

\begin{table}[]
\caption{Number of trades over the 10-year learning simulation period averaged between the four highest performing simulations, and also the calculated average number of trades per day among the top four performers.}
\label{table:numberoftrades}
\begin{tabular}{|l|l|l|}
\hline
\textbf{Average among the top four performers}                                     & \textbf{Value} & \textbf{$\mathbf{\sigma}$} \\ \hline
Number of trades over 10 years & 48,078         & 12,641 \\ \hline
Average number of trades per day & 13.16          & --- \\ \hline
\end{tabular}
\end{table}

\subsubsection{Heuristics Performance}\label{section:heuristics-performance}

The best possible performance of the two heuristics is summarized in Table~\ref{table:heuristicperformance}. As we can see, the performance of the trend reversal heuristic is much better than of the trend continuation one. Also, we observe smaller average number of trades for the trend continuation heuristic. It is also worth mentioning that, as it is seen in Table~\ref{table:appendixA}, among all the 21 cases with positive final balances, only three are produced by the trend continuation heuristic, and all these cases are among bottom five performers with positive balances. 

\begin{table}[]
\caption{Summary of best possible performance of the two heuristic prototypes.}
\label{table:heuristicperformance}
\begin{tabular}{|l|l|l|l|}
\hline
\textbf{\begin{tabular}[c]{@{}l@{}}Heuristic\\ Prototype\end{tabular}} & \textbf{\begin{tabular}[c]{@{}l@{}}Average 10-year\\ Balance (pips)\end{tabular}} & \textbf{\begin{tabular}[c]{@{}l@{}}Avg. Number of \\ Trades over 10 years\end{tabular}} & \textbf{$\mathbf{\sigma}$} \\ \hline
$h_1$                                                                    & -10,574                                                                           & 15,101                                                                                  & 13,997         \\ \hline
$h_2$                                                                    & 186,869                                                                           & 19,007                                                                                  & 18,399         \\ \hline
\end{tabular}
\end{table}

\subsubsection{Optimal TP/SL Settings}

Table~\ref{table:optimal-ttsl} is the culmination of this research, delivering the set of optimal trading settings that allows to achieve the best profit results.

\begin{table}[H]
\caption{Optimal TP/SL parameters for the top 7 most profitable simulation cases.}
\label{table:optimal-ttsl}
\begin{tabular}{|l|l|l|l|}
\hline
\textbf{Instrument} & \textbf{Period} & \textbf{\begin{tabular}[c]{@{}l@{}}Take Profit\\ (pips)\end{tabular}} & \textbf{\begin{tabular}[c]{@{}l@{}}Stop Loss\\ (pips)\end{tabular}} \\ \hline
EUR/JPY             & H1              & 3                                                                     & 14                                                                  \\ \hline
EUR/USD             & H1              & 3                                                                     & 11                                                                  \\ \hline
EUR/JPY             & H2              & 3                                                                     & 20                                                                  \\ \hline
USD/JPY             & H1              & 3                                                                     & 9                                                                   \\ \hline
EUR/USD             & H2              & 3                                                                     & 16                                                                  \\ \hline
EUR/JPY             & H4              & 3                                                                     & 28                                                                  \\ \hline
USD/JPY             & H2              & 3                                                                     & 13                                                                  \\ \hline
\end{tabular}
\end{table}

\section{Discussion}
\smartfx is purposefully designed to be broker-agnostic, which allows to develop trading setups that are universally applicable to diverse trading environments. However, this decision comes at a cost of the necessity to specify a fixed spread value, whereas many Forex brokers use floating spread policy. As a result, the user of \smartfx must specify the spread that is large enough. Although the static spread value does not compromise the quality of the result produced by \smartfx, the parameter must be chosen carefully in order to avoid overly optimistic scenarios.

\section{Conclusion}
Market trading optimization is a popular application of Artificial Intelligence (AI), delivering endless opportunities for research and profound insights. The Forex market is often being chosen as a target environment because it allows unlimited spread-based day trading without additional fees, long market hours, and easy entrance requirements. In this research, we demonstrated that popular technical indicators may be ambiguous and less optimal for automated trading. We developed \smartfx, an automated constraint-based machine learning tool that determines optimal parameters for given heuristic prototypes.

We evaluated 10 years worth of market data for 3 popular Forex instruments using two simple price-action heuristic prototypes, along with 6 different OHLC granularities of the input data. The results clearly demonstrated that the trade reversal price-action heuristic is significantly more profitable than the trade continuation, and capable of delivering up to 118 pips of profit per day on average. We determined exact trading parameters for the optimal trade, and thus achieved the goal of this research: to deliver practically working, specific and evidence-based optimal trade parameters for price-action Forex trading. We make \smartfx available under an open source license for everyone to use, evaluate, and build upon.

\clearpage
\bibliographystyle{IEEEtran}

\begin{thebibliography}{10}
\bibitem{britanicaforex}
D.~Hudson, ``Foreign exchange market,'' Encyclopædia Britannica, 2017,
  accessed: 2019-12-07.

\bibitem{thompson2017time}
G.~F. Thompson, ``Time, trading and algorithms in financial sector security,''
  \emph{New Political Economy}, vol.~22, no.~1, pp. 1--11, 2017.

\bibitem{yazdi2013technical}
S.~H.~M. Yazdi and Z.~H. Lashkari, ``Technical analysis of forex by macd
  indicator,'' \emph{International Journal of Humanities and Management
  Sciences (IJHMS)}, vol.~1, no.~2, pp. 159--165, 2013.

\bibitem{5358946}
Z.~{Liu} and D.~{Xiao}, ``An automated trading system with multi-indicator
  fusion based on d-s evidence theory in forex market,'' in \emph{2009 Sixth
  International Conference on Fuzzy Systems and Knowledge Discovery}, vol.~3,
  Aug 2009, pp. 239--243.

\bibitem{krishnan2009impact}
R.~Krishnan and S.~S. Menon, ``Impact of currency pairs, time frames and
  technical indicators on trading profit in forex spot market.''
  \emph{International journal of Business insights \& Transformation}, vol.~2,
  no.~2, 2009.

\bibitem{yong2015technical}
Y.~L. Yong, D.~C. Ngo, and Y.~Lee, ``Technical indicators for forex
  forecasting: a preliminary study,'' in \emph{International Conference in
  Swarm Intelligence}.\hskip 1em plus 0.5em minus 0.4em\relax Springer, 2015,
  pp. 87--97.

\bibitem{ozturk2016heuristic}
M.~Ozturk, I.~H. Toroslu, and G.~Fidan, ``Heuristic based trading system on
  forex data using technical indicator rules,'' \emph{Applied Soft Computing},
  vol.~43, pp. 170--186, 2016.

\bibitem{yazdi2012technical}
S.~H.~M. Yazdi and Z.~H. LASHKARI, ``Technical analysis of forex by parabolic
  sar indicator,'' in \emph{International Islamic Accounting and Finance
  Conference}, 2012.

\bibitem{ni2009exchange}
H.~Ni and H.~Yin, ``Exchange rate prediction using hybrid neural networks and
  trading indicators,'' \emph{Neurocomputing}, vol.~72, no. 13-15, pp.
  2815--2823, 2009.

\bibitem{rosenstreich2005forex}
P.~Rosenstreich, \emph{Forex Revolution: An Insider's Guide to the Real World
  of Foreign Exchange Trading}.\hskip 1em plus 0.5em minus 0.4em\relax FT
  Press, 2005.

\bibitem{guides2017personality}
T.~S. Guides, ``Personality strengths and weaknesses in forex trading,'' 2017.

\bibitem{de2012comparative}
R.~F. de~Brito and A.~L. Oliveira, ``Comparative study of forex trading systems
  built with svr+ ghsom and genetic algorithms optimization of technical
  indicators,'' in \emph{2012 IEEE 24th International Conference on Tools with
  Artificial Intelligence}, vol.~1.\hskip 1em plus 0.5em minus 0.4em\relax
  IEEE, 2012, pp. 351--358.

\bibitem{8096360}
T.~{Ploysuwan} and R.~{Chaisricharoen}, ``Gaussian process kernel crossover for
  automated forex trading system,'' in \emph{2017 14th International Conference
  on Electrical Engineering/Electronics, Computer, Telecommunications and
  Information Technology (ECTI-CON)}, June 2017, pp. 802--805.

\bibitem{wilson2010interday}
G.~Wilson and W.~Banzhaf, ``Interday foreign exchange trading using linear
  genetic programming,'' in \emph{Proceedings of the 12th annual conference on
  Genetic and evolutionary computation}.\hskip 1em plus 0.5em minus 0.4em\relax
  ACM, 2010, pp. 1139--1146.

\bibitem{pinto2012strategic}
R.~M.~C. Pinto and J.~C.~M. Silva, ``Strategic methods for automated trading in
  forex,'' in \emph{2012 12th International Conference on Intelligent Systems
  Design and Applications (ISDA)}.\hskip 1em plus 0.5em minus 0.4em\relax IEEE,
  2012, pp. 34--39.

\bibitem{vezeris2019adturtle}
D.~Vezeris, I.~Karkanis, and T.~Kyrgos, ``Adturtle: An advanced turtle trading
  system,'' \emph{Journal of Risk and Financial Management}, vol.~12, no.~2,
  p.~96, 2019.

\bibitem{schoen2011automated}
J.~E. Schoen, ``Automated trading system,'' Jan.~4 2011, uS Patent 7,865,421.

\bibitem{chait2007foreign}
J.~Chait, ``Foreign exchange trading platform,'' Feb.~22 2007, uS Patent App.
  11/451,731.

\bibitem{peterson2005foreign}
A.~Peterson, L.~Miller, I.~Eaglesfield, S.~Hubble, M.~LeGelebart, T.~Sablic,
  R.~Sagurton, E.~Theodorou, W.~Wah, A.~Zhou \emph{et~al.}, ``Foreign exchange
  trading interface,'' Mar.~17 2005, uS Patent App. 10/884,111.

\bibitem{ogg2005automated}
D.~Ogg, W.~Lamartin, and M.~Smith, ``Automated system for routing orders for
  foreign exchange transactions,'' Nov.~3 2005, uS Patent App. 10/834,812.

\bibitem{owens2005automated}
J.~Owens and C.~Awtry, ``Automated trading system and software for financial
  markets,'' Dec.~22 2005, uS Patent App. 11/156,428.

\bibitem{983110}
A.~{Abraham} and M.~U. {Chowdhury}, ``An intelligent forex monitoring system,''
  in \emph{2001 International Conferences on Info-Tech and Info-Net.
  Proceedings (Cat. No.01EX479)}, vol.~3, Oct 2001, pp. 523--528 vol.3.

\bibitem{abraham2001intelligent}
A.~Abraham and M.~U. Chowdhury, ``An intelligent forex monitoring system,'' in
  \emph{2001 International Conferences on Info-Tech and Info-Net. Proceedings
  (Cat. No. 01EX479)}, vol.~3.\hskip 1em plus 0.5em minus 0.4em\relax IEEE,
  2001, pp. 523--528.

\bibitem{czekalski2015ann}
P.~Czekalski, M.~Niezabitowski, and R.~Styblinski, ``Ann for forex forecasting
  and trading,'' in \emph{2015 20th International Conference on Control Systems
  and Computer Science}.\hskip 1em plus 0.5em minus 0.4em\relax IEEE, 2015, pp.
  322--328.

\bibitem{lee2014hidden}
Y.~Lee, L.~C.~O. Tiong, and D.~C.~L. Ngo, ``Hidden markov models for forex
  trends prediction,'' in \emph{2014 International Conference on Information
  Science \& Applications (ICISA)}.\hskip 1em plus 0.5em minus 0.4em\relax
  IEEE, 2014, pp. 1--4.

\bibitem{barbosa2010multi}
R.~P. Barbosa and O.~Belo, ``Multi-agent forex trading system,'' in \emph{Agent
  and multi-agent technology for internet and enterprise systems}.\hskip 1em
  plus 0.5em minus 0.4em\relax Springer, 2010, pp. 91--118.

\bibitem{shahbazi2016forex}
N.~Shahbazi, M.~Memarzadeh, and J.~Gryz, ``Forex market prediction using narx
  neural network with bagging,'' in \emph{MATEC Web of Conferences},
  vol.~68.\hskip 1em plus 0.5em minus 0.4em\relax EDP Sciences, 2016, p. 19001.

\bibitem{macerinskiene2015evidence}
I.~Macerinskiene and A.~Balciunas, ``The evidence of social responsibility in
  foreign exchange brokers’ activities,'' \emph{Procedia-Social and
  Behavioral Sciences}, vol. 213, pp. 552--556, 2015.

\end{thebibliography}



\clearpage
\appendix

\subsection{Resulting Balances}

Table~\ref{table:appendixA} shows the full list of ending balances for each of the learning simulation cases. The column Reverse represents the heuristic: if it is $FALSE$, $h_1$ is used, otherwise, $h_2$ is used. Depending on the instrument, the multiplier for determining the number of pips in the third column is either $10^2$ or $10^4$.

\begin{table}[H]
\tiny
\caption{Resulting balances for all the simulations. If the value in the Reverse column is $FALSE$, the the heuristic $h_1$ is used, otherwise, the heuristic $h_2$ is used. The data is sorted in a descending order by the value of the third column.} 
\label{table:appendixA}
\centering
\begin{tabular}{|l|l|l|l|}
\hline
\textbf{Instrument} & \textbf{Balance} & \textbf{Balance in Pips} & \textbf{Reversal} \\ \hline
EURJPY-H1           & 4321.923         & 432192.3                  & TRUE              \\ \hline
EURUSD-H1           & 34.86112         & 348611.2                  & TRUE              \\ \hline
EURJPY-H2           & 3020.2202        & 302022.02                 & TRUE              \\ \hline
USDJPY-H1           & 2941.217         & 294121.7                  & TRUE              \\ \hline
EURUSD-H2           & 23.923656        & 239236.56                 & TRUE              \\ \hline
EURJPY-H4           & 2191.7815        & 219178.15                 & TRUE              \\ \hline
USDJPY-H2           & 1984.1091        & 198410.91                 & TRUE              \\ \hline
EURUSD-H4           & 17.544798        & 175447.98                 & TRUE              \\ \hline
EURJPY-H8           & 1635.9933        & 163599.33                 & TRUE              \\ \hline
EURJPY-H12          & 1460.575         & 146057.5                  & TRUE              \\ \hline
USDJPY-H4           & 1441.6462        & 144164.62                 & TRUE              \\ \hline
EURUSD-H8           & 12.996401        & 129964.01                 & TRUE              \\ \hline
EURUSD-H12          & 12.005339        & 120053.39                 & TRUE              \\ \hline
USDJPY-H8           & 1074.1615        & 107416.15                 & TRUE              \\ \hline
EURJPY-D1           & 963.3696         & 96336.96                  & TRUE              \\ \hline
USDJPY-H12          & 924.0237         & 92402.37                  & TRUE              \\ \hline
USDJPY-H1           & 872.2552         & 87225.52                  & FALSE             \\ \hline
EURUSD-D1           & 8.5002575        & 85002.575                 & TRUE              \\ \hline
USDJPY-D1           & 694.33014        & 69433.014                 & TRUE              \\ \hline
EURUSD-H1           & 5.1108055        & 51108.055                 & FALSE             \\ \hline
USDJPY-H2           & 183.90222        & 18390.222                 & FALSE             \\ \hline
EURJPY-H1           & -14.003174       & -1400.3174                & FALSE             \\ \hline
EURJPY-D1           & -78.99794        & -7899.794                 & FALSE             \\ \hline
EURUSD-D1           & -0.8153805       & -8153.805                 & FALSE             \\ \hline
USDJPY-D1           & -81.747986       & -8174.7986                & FALSE             \\ \hline
EURJPY-H12          & -148.27484       & -14827.484                & FALSE             \\ \hline
EURUSD-H12          & -1.4840896       & -14840.896                & FALSE             \\ \hline
USDJPY-H12          & -150.52483       & -15052.483                & FALSE             \\ \hline
EURUSD-H8           & -2.1492684       & -21492.684                & FALSE             \\ \hline
EURJPY-H8           & -215.3318        & -21533.18                 & FALSE             \\ \hline
USDJPY-H8           & -218.21217       & -21821.217                & FALSE             \\ \hline
USDJPY-H4           & -257.88776       & -25788.776                & FALSE             \\ \hline
EURUSD-H2           & -2.9397478       & -29397.478                & FALSE             \\ \hline
EURUSD-H4           & -3.9952888       & -39952.888                & FALSE             \\ \hline
EURJPY-H4           & -413.3832        & -41338.32                 & FALSE             \\ \hline
EURJPY-H2           & -753.8009        & -75380.09                 & FALSE             \\ \hline
\end{tabular}
\end{table}

\subsection{Stop Loss and Take Profit Configurations}

Table~\ref{table:appendixB} contains the most optimal take profit and stop loss parameters for all simulation cases that ended up with positive balances. Recording the cases with negative balances was not necessary because, by all means, these cases cannot be defined as successful, and therefore the values of most optimal take profit and stop loss parameters are never practically useful.

\begin{table}[]
\scriptsize
\caption{Most profitable take profit (TP) and stop loss (SL) settings for all learning simulation cases that ended up with a positive balance.}
\label{table:appendixB}
\centering
\begin{tabular}{|l|l|l|l|}
\hline
\textbf{Simulation} & \textbf{SL} & \textbf{TP} & \textbf{Reverse} \\ \hline
EURJPY-H1           & 0.1400      & 0.0300      & TRUE             \\ \hline
EURUSD-H1           & 0.0011      & 0.0003      & TRUE             \\ \hline
EURJPY-H2           & 0.2000      & 0.0300      & TRUE             \\ \hline
USDJPY-H1           & 0.0900      & 0.0300      & TRUE             \\ \hline
EURUSD-H2           & 0.0016      & 0.0003      & TRUE             \\ \hline
EURJPY-H4           & 0.2800      & 0.0300      & TRUE             \\ \hline
USDJPY-H2           & 0.1300      & 0.0300      & TRUE             \\ \hline
EURUSD-H4           & 0.0024      & 0.0003      & TRUE             \\ \hline
EURJPY-H8           & 0.4500      & 0.0300      & TRUE             \\ \hline
EURJPY-H12          & 0.4800      & 0.0300      & TRUE             \\ \hline
USDJPY-H4           & 0.1800      & 0.0300      & TRUE             \\ \hline
EURUSD-H8           & 0.0038      & 0.0003      & TRUE             \\ \hline
EURUSD-H12          & 0.0047      & 0.0003      & TRUE             \\ \hline
USDJPY-H8           & 0.2700      & 0.0300      & TRUE             \\ \hline
EURJPY-D1           & 0.5000      & 0.0300      & TRUE             \\ \hline
USDJPY-H12          & 0.3500      & 0.0300      & TRUE             \\ \hline
USDJPY-H1           & 0.0300      & 0.5000      & FALSE            \\ \hline
EURUSD-D1           & 0.0050      & 0.0003      & TRUE             \\ \hline
USDJPY-D1           & 0.5000      & 0.0300      & TRUE             \\ \hline
EURUSD-H1           & 0.0004      & 0.0050      & FALSE            \\ \hline
USDJPY-H2           & 0.0600      & 0.5000      & FALSE            \\ \hline
\end{tabular}
\end{table}

\subsection{Number of Trades}\label{appendix:numberoftrades}
Table~\ref{table:appendixC} contains total number of trades for ten years, and also a calculated value of an average number of trades expected per one day, which is simply the number in the second column divided by $7 \cdot 365 + 3 \cdot 366$ An important note about the days when the market is closed: normally, during these days there are no trades, so, depending on what the result is used for, the number in the second column may be instead divided by $(7 \cdot 365 + 3 \cdot 366) \times {5 \over 7}$.

\begin{table}[]
\scriptsize
\caption{Number of trades for each of the simulation learning cases, including the calculated average trades per day value.}
\label{table:appendixC}
\centering
\begin{tabular}{|l|l|l|}
\hline
\textbf{Simulation} & \textbf{Number of Trades} & \textbf{Average Trades per Day} \\ \hline
EURJPY-H1           & 56078                     & 15.35                           \\ \hline
EURUSD-H1           & 53299                     & 14.59                           \\ \hline
EURJPY-H2           & 29205                     & 7.99                            \\ \hline
USDJPY-H1           & 53729                     & 14.71                           \\ \hline
EURUSD-H2           & 27823                     & 7.62                            \\ \hline
EURJPY-H4           & 15318                     & 4.19                            \\ \hline
USDJPY-H2           & 28013                     & 7.67                            \\ \hline
EURUSD-H4           & 14670                     & 4.02                            \\ \hline
EURJPY-H8           & 7977                      & 2.18                            \\ \hline
EURJPY-H12          & 5594                      & 1.53                            \\ \hline
USDJPY-H4           & 14838                     & 4.06                            \\ \hline
EURUSD-H8           & 7821                      & 2.14                            \\ \hline
EURUSD-H12          & 5515                      & 1.51                            \\ \hline
USDJPY-H8           & 7786                      & 2.13                            \\ \hline
EURJPY-D1           & 3062                      & 0.84                            \\ \hline
USDJPY-H12          & 5425                      & 1.49                            \\ \hline
USDJPY-H1           & 48709                     & 13.33                           \\ \hline
EURUSD-D1           & 3007                      & 0.82                            \\ \hline
USDJPY-D1           & 2967                      & 0.81                            \\ \hline
EURUSD-H1           & 43709                     & 11.97                           \\ \hline
USDJPY-H2           & 18504                     & 5.07                            \\ \hline
EURJPY-H1           & 37402                     & 10.24                           \\ \hline
EURJPY-D1           & 3123                      & 0.85                            \\ \hline
EURUSD-D1           & 3126                      & 0.86                            \\ \hline
USDJPY-D1           & 3118                      & 0.85                            \\ \hline
EURJPY-H12          & 5714                      & 1.56                            \\ \hline
EURUSD-H12          & 5723                      & 1.57                            \\ \hline
USDJPY-H12          & 5711                      & 1.56                            \\ \hline
EURUSD-H8           & 8314                      & 2.28                            \\ \hline
EURJPY-H8           & 8314                      & 2.28                            \\ \hline
USDJPY-H8           & 8300                      & 2.27                            \\ \hline
USDJPY-H4           & 9440                      & 2.58                            \\ \hline
EURUSD-H2           & 13397                     & 3.67                            \\ \hline
EURUSD-H4           & 16087                     & 4.40                            \\ \hline
EURJPY-H4           & 16096                     & 4.41                            \\ \hline
EURJPY-H2           & 17044                     & 4.67                            \\ \hline
\end{tabular}
\end{table}

\end{document}